\documentclass[aps,prl,twocolumn,a4paper,longbibliography,superscriptaddress]{revtex4-1}
\usepackage{graphicx}
\usepackage{amsmath}

\begin{document}
\title{\bf Electron correlations in Mn$_x$Ga$_{1-x}$As as seen by resonant electron spectroscopy and dynamical mean field theory}

\author{I. Di Marco}
\email{igor.dimarco@physics.uu.se}
\author{P. Thunstr\"om}
\affiliation{Department of Physics and Astronomy, Uppsala University, Box 516, SE-75120, Uppsala, Sweden}
\author{M. I. Katsnelson}
\affiliation{Radboud University Nijmegen, Institute for Molecules
and Materials, NL-6525 AJ Nijmegen, The Netherlands}
\author{J. Sadowski}
\affiliation{MAX-lab, Lund University, SE-22100 Lund, Sweden and \\
 Institute of Physics, Polish Academy of Sciences, al. Lotnikow 32/46, 02-668 Warszawa Poland}
\author{K. Karlsson}
\affiliation{Department of Life Sciences, H\"ogskolan i Sk\"ovde, SE-54128 Sk\"ovde, Sweden}
\author{S. Leb\`egue}
\affiliation{
Laboratoire de Cristallographie, R\'esonance Magn\'etique et Mod\'elisations (CRM2, UMR CNRS 7036)
 Institut Jean Barriol, Universit\'e de Lorraine,
 BP 239, Boulevard des Aiguillettes
 54506 Vandoeuvre-l\`es-Nancy,France}
\author{J. Kanski}
\affiliation{Chalmers, Department of Applied Physics, SE-41296 Gothenburg, Sweden }
\author{O. Eriksson}
\affiliation{Department of Physics and Astronomy, Uppsala University, Box 516, SE-75120, Uppsala, Sweden}

\date{\today }

\maketitle

{\bf 
After two decades from the discovery of ferromagnetism in Mn-doped GaAs, its origin is still debated, and many doubts are related to the electronic structure. Here we report an experimental and theoretical study of the valence electron spectrum of Mn-doped GaAs. The experimental data are obtained through the differences between off- and on-resonance photo-emission data. The theoretical spectrum is calculated by means of a combination of density-functional theory in the local density approximation and dynamical mean-field theory (LDA+DMFT), using exact diagonalisation as impurity solver. Theory is found to accurately reproduce measured data, and illustrates the importance of correlation effects. Our results demonstrate that the Mn states extend over a broad range of energy, including the top of the valence band, and that no impurity band splits off from the valence band edge, while the induced holes seem located primarily around the Mn impurity.
}

The field of diluted magnetic semiconductors (DMS) \cite{ohno98science281:951,dietl00science287:5455} has for the past decade broadened to include several aspects of materials science, e.g. materials synthesis in bulk and thin film form, structural and magnetic characterization, photo-electron spectroscopy for characterization of the electronic structure, and theoretical approaches based on model Hamiltonians or first principles calculations. Several reviews have lately been authored, e.g. Refs. \onlinecite{zunger_book_1986,jungwirth06rmp78:809,pulizzi08naturemilestone,sato10rmp82:1633,dietl10nat9:965}, presenting the evolution of the field starting from the first reports on magnetic impurities in semiconductors \cite{cochrane74prb9:3013,story86prl56:777}. A milestone event was the report on carrier-mediated ferromagnetism of Mn-doped InAs and GaAs \cite{ohno98science281:951}, where the Mn atoms both act as acceptors and magnetic impurities. 

In practical applications of spintronics, an ordering temperature above room temperature is needed, and a few systems were predicted \cite{dietl00science287:5455} to be promising from that point of view, namely Mn-doped ZnO and Mn-doped GaN. Unfortunately,  it has proven very difficult to determine their ordering temperatures  experimentally, and values ranging from close to 1000 K to just above 0 K have been reported. This vast spread of values can possibly be linked to difficulties in synthesizing samples with a homogeneous distribution of substitutional Mn atoms in the host lattice. Mn-doped GaAs offers a considerably more robust situation from a synthesis point of view, and substitutional Mn doping  at the Ga sites can reach concentrations up to around 15\%~\cite{chiba07apl90:122503,chen09apl95:182505}, as estimated from saturation magnetization data. The magnetic ordering temperature ($\rm T_C$) depends on the detailed preparation procedures and evaluation methods~\cite{chen09apl95:182505,wang05aip_cp772:333,wang08apl93:132103}, the highest reported value being between 188 K~\cite{nemec13natcomm4:1422} and about 200 K~\cite{novak08prl101:077201}, inferred respectively from the measured magnetization and from the maximum in the resistivity curve (the latter method is known to overestimate the transition temperature~\cite{chiba07apl90:122503}).

On the theory side, it has been suggested~\cite{dietl00science287:5455} that the ferromagnetic interaction in Mn-doped GaAs is described by Vonsovsky-Zener's p-d exchange model~\cite{vonsovsky46zetf16:981,zener51pr81:440}. The main assumptions of this model are that the Mn-3d electrons are localized, and that the divalent Mn impurities can induce itinerant spin-carriers (holes) in the valence band of the host material~\cite{dietl_book,jungwirth06rmp78:809}. These holes couple anti-ferromagnetically to the localized moments, and mediate an effective long-range exchange interaction between the moments themselves~\cite{dietl00science287:5455, dietl_book}. The picture of long-ranged interactions has, however, been challenged by calculations based on first principles theory, albeit without correlation effects taken into account. These works suggest that the spin carrying holes are more bound to the Mn impurity, and arise from a narrow impurity band of mixed spd character~\cite{sato10rmp82:1633,jungwirth06rmp78:809}. As a result a much more short ranged interaction is observed, with magnetic percolation as a key ingredient, and a decent agreement is found between calculated and observed ordering temperatures~\cite{bergqvist04prl93:137202,sato04prb70:201202}. This situation is better described by the double-exchange model, which was originally introduced for mixed valence dense ferromagnets~\cite{zener51pr82:403} and later on adapted to DMS~\cite{nagaev_book,sanvito01prb63:165206,jungwirth06rmp78:809,sato10rmp82:1633}. In the double exchange model it is not crucial if the main character of the carriers is 3d~\cite{dietl10nat9:965,sato10rmp82:1633} or 4p~\cite{krstajic03epl61:235,jungwirth06rmp78:809} as this does not affect either the basic physics (dependence of the effective hopping parameter on the magnetic configuration) nor the form of the effective magnetic Hamiltonian~\cite{auslender82ssc44:387,auslender82tmp51:601}. Instead, the essential condition is that the band crossing the Fermi level has a width which is smaller than the effective exchange energy between localized moments and spin-carrying holes~\cite{jungwirth06rmp78:809}. This is opposite to the situation of the aforementioned p-d exchange model, where the exchange energy is smaller than the bandwidth of the carriers. In this view double exchange and p-d exchange are both due to the hybridization between conduction and magnetic electrons, and are respectively associated to the limit of narrow and broad bands~\cite{nagaev_book,jungwirth06rmp78:809,katsnelson08rmp80:315}.

Apart from first-principles studies, a significant contribution from the double exchange mechanism in Mn-doped GaAs is also suggested in a recent experimental study~\cite{ohya10prl104:167204}.  Here an impurity level was conferred to lie in the band gap, which was intersected by the Fermi level. The experimental work of Dodobrowolska {\itshape{et al.}}~\cite{dobrowolska12nm11:444} comes to similar conclusions. In particular it was reported that the Fermi level lies above the top of the valence band, and although a direct measurement of an impurity band was lacking, the authors interpreted their data in terms of the existence of such a feature.
Spectroscopic investigations, based on hard x-ray angle resolved photo-emission~\cite{gray12natmat11:957}, do not find such an impurity level. This work instead proposed that both p-d as well as double exchange are active in Mn-doped GaAs, as suggested in Ref.~\onlinecite{sato10rmp82:1633}.

One must conclude that there is a lack of consensus regarding the mechanisms behind magnetism in Mn-doped GaAs and possibly DMS systems in general. The shortcoming in understanding magnetism in these materials  is coupled to the lack of a precise knowledge of their electronic structure. A weak point in previous first-principles studies through density functional theory, is that the calculations are based on a mean-field like description of the electronic structure, using the local density approximation (LDA) or the generalized gradient approximation (GGA). The main criticism to this theory is that a mean-field like description may not be sufficiently accurate for describing the Mn-3d states, and that correlation effects due to the strong Coulomb interaction must be considered~\cite{dietl10nat9:965}. 
Indeed, calculations including correlation effects have been made for simplified models with reduced degrees of freedom, using several adjustable parameters~\cite{chaco03prb68:233310,majidi06prb74:115205,popescu06prb73:075206,popescu07prb76:085206}. While these models have been important to identify the different competing mechanisms in various parameter regimes, their applicability to Mn-doped GaAs cannot be inferred from a comparison with an accurate experimental photo-emission spectrum. Among the published studies, the only materials dependent theory including electron-electron interaction directly is the mean-field (static) LDA+U method~\cite{park00pb281:703,sanyal03prb68:205210,shick04prb69:125207,sandratskii04prb69:195203,sato10rmp82:1633}, but as we show below, this method is inaccurate when it comes to reproduce the measured electronic structure of Mn-doped GaAs. 

The discussion above indicates that a thorough analysis of the electronic structure of Mn-doped GaAs is highly needed, combining experimental data with accurate theory, to clarify the role of electron correlations and their impact on magnetism. Recent studies regarding the coupling of the electronic structure to the nature of the magnetic interactions~\cite{dobrowolska12nm11:444}, and the resulting debate on their validity~\cite{edmonds_arxiv,dobrowolska_arxiv}, further highlight this need.
The purpose of this article is to provide such an analysis, and we present here the spectral properties of Mn-doped GaAs as given by differences between off- and on-resonance photo-emission data for 0.1, 1 and 6\% Mn doping. The experimental results are compared with electronic structure calculations based on a combination of density-functional theory and dynamical mean-field theory (LDA+DMFT) \cite{kotliar06rmp78:865,held07ap56:829,katsnelson08rmp80:315}. Both theory and experiment are focused on the high-temperature paramagnetic phase, as described in the Method section.

\section{Results}
\subsection{Correlated Mn-3d states}
\begin{figure}
\includegraphics[width=8.5cm]{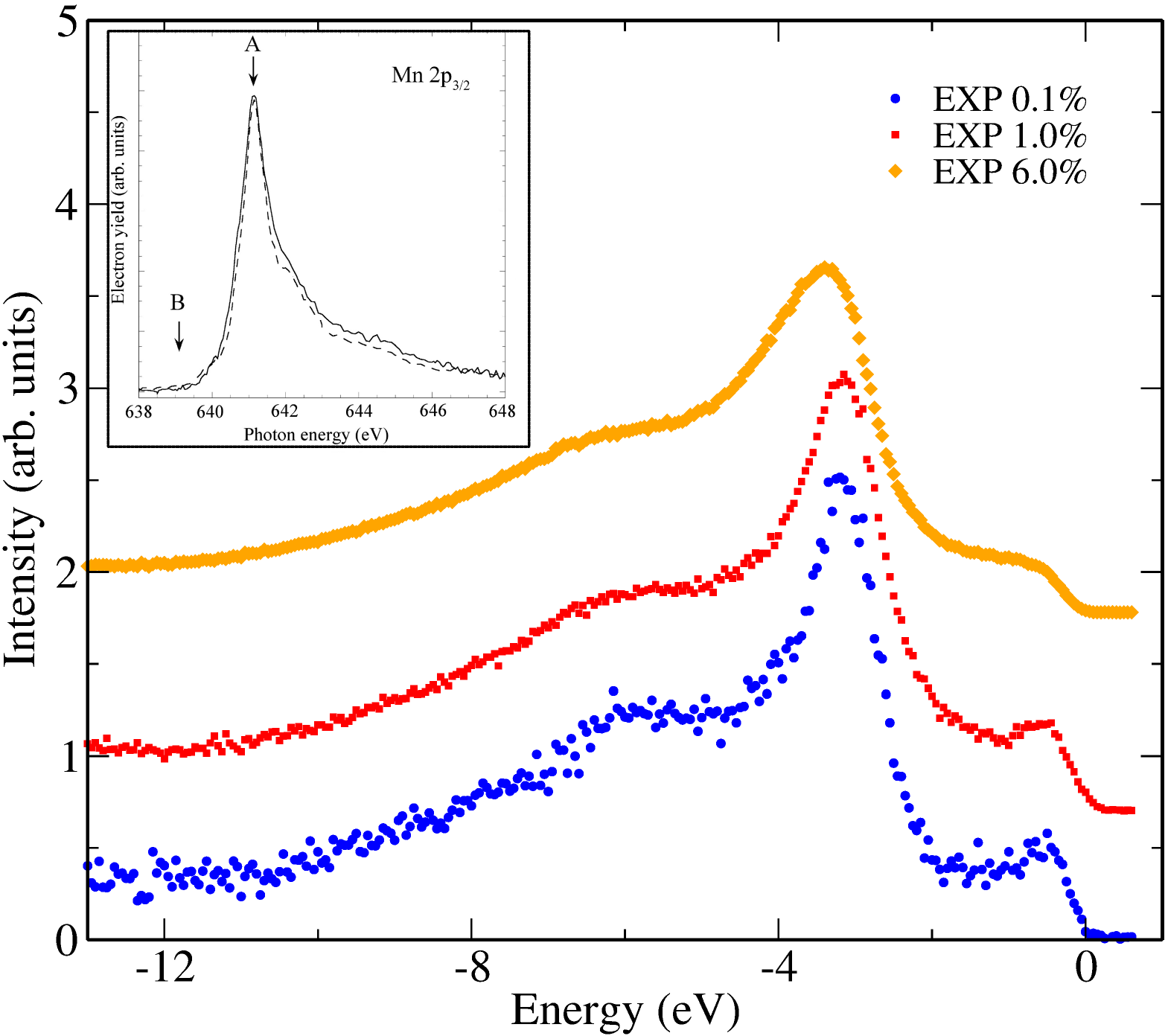}
\caption{{\bf{Resonant photo-emission spectra of Mn-3d states.}} Data for various samples of Mn-doped GaAs at different impurities concentration. The 0.1\% and 1\% spectra are shifted in energy to align with the 6\% spectrum at the valence band maximum. In the last case the valence band maximum coincides with the Fermi level. In the inset the Mn 2p$_{3/2}$ XAS spectra with 0.3\% and 6\% Mn (dashed and solid lines, respectively) are shown. "A" and "B" indicate the two photon energies used for extracting the resonant photo-emission spectra.}
\label{fig:Mn_exp_spectrum}
\end{figure}

\begin{figure}
\includegraphics[width=8.5cm]{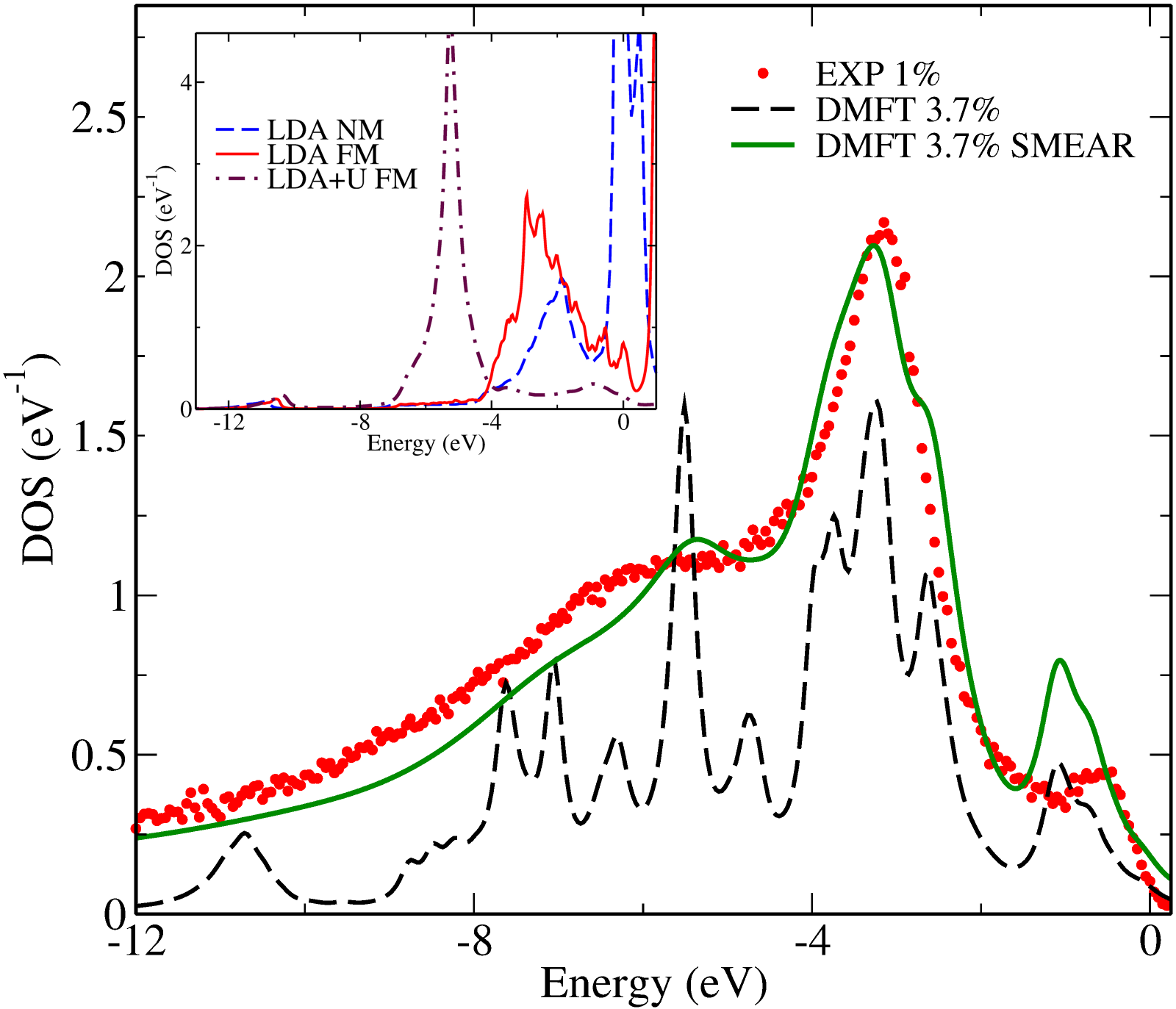}
\caption{{\bf{Mn-3d PDOS from LDA+DMFT.}} Resonant photo-emission spectrum of Mn-3d states in Mn-doped GaAs at 1\% concentration compared to the Mn-3d projected density of states from LDA+DMFT, with and without Lorentzian smearing. The energy dependent width of the latter was chosen as $\tau(\epsilon)=0.04 \text{ eV}^{-1} \cdot \epsilon^2 + 0.04 \text{ eV}$, in order to allow a better comparison of the spectral weight with the photo-emission signal. The Fermi level is at zero energy, and no artificial shifts have been applied between theoretical and experimental curves. In the inset the ferromagnetic (FM) and non-magnetic (NM) LSDA results are shown, together with the LDA+U results.}
\label{fig:Mn3d_spectrum}
\end{figure}
 
In Fig.~\ref{fig:Mn_exp_spectrum} the experimental data for the Mn-3d one-particle excitation spectrum are reported for samples with a Mn concentration of {0.1\%}, {1\%}, and {6\%}. The curves were extracted by taking differences between off- and on-resonance photon energies for the Mn-3d cross section. In Fig.~\ref{fig:Mn3d_spectrum} the experimental spectrum for the sample with {1\%} Mn concentration is directly compared with the LDA+DMFT projected density of states (PDOS) of the Mn-3d orbitals, for {3.7\%} Mn concentration. For such concentrations of Mn, the direct wave-function overlap between different Mn atoms is vanishingly small, and it is expected that the spectral properties are rather insensitive to smaller modifications of the Mn concentration (as Fig.~\ref{fig:Mn_exp_spectrum} also shows). Comparison with the more diluted experimental spectrum is motivated by the fact that at high Mn concentrations the spectrum is broadened due to a statistical (i.e. non-uniform) Mn distribution. For Mn atoms occupying Ga sites in the zincblende structure the probability for finding two nearest neighbor sites occupied is 30\%~\cite{ulfat} at 3\% Mn concentration. Note that we in Fig.~\ref{fig:Mn3d_spectrum} show theoretical data, directly as obtained from the DMFT calculation, as well as theory which has been broadened due to life-time effects. The broadening considered was obtained from Fermi-Liquid theory, in which the life-time decreases quadratically with the binding energy~\cite{echenique00cp251:1}. As is clear from Fig.~\ref{fig:Mn3d_spectrum}, both experiment and theory position the Mn-3d states primarily in the valence band of the GaAs host, with the Mn-3d states extending over a broad range of energies. 
The experimental spectrum has a main peak at $\sim$ 3.2 eV binding energy, an observation that is captured by theory. Furthermore, the observed broad shoulder, which is located between 4 and 8 eV binding energy, is also well described in the LDA+DMFT excitation spectrum. Overall, the agreement between theory and experiment must be judged as being good.
Only in the closest region to the Fermi level can one notice some discrepancies between theory and experiment, where the small peak at $\sim$ 1 eV binding energy has a slightly too large intensity in the calculations, when compared to observations.

\subsection{Host valence band and induced holes}
In the inset of Fig.~\ref{fig:Mn3d_spectrum} the Mn-3d PDOS for spin-polarized and unpolarized LSDA are reported, and we note that the agreement with the experimental data is less than satisfactory. Focusing on the ferromagnetic LSDA solution, the spectral weight is almost entirely contained in the region with excitation energies below 4 eV, and while the position of the main peak is almost correct, the large shoulder at higher binding energy is totally absent. If electron-electron repulsion is treated in a static Hartree-Fock fashion, as for the LDA+U calculations shown in the same inset, the agreement with the experiment does not improve much. The main peak is moved to higher binding energies, its exact position depending on the value of $U$~\cite{park00pb281:703,sanyal03prb68:205210,shick04prb69:125207,sandratskii04prb69:195203}, but it is not possible to describe the broad distribution of spectral weight at binding energies between 4 and 8 eV. One must conclude that overall the electronic structure of Mn-doped GaAs is not accurately described by LDA or LDA+U theory, and that dynamical correlation effects play a main role. The experimental results of Ref.~\onlinecite{gray12natmat11:957} were in the same study compared to a theory based on LDA and interfaced with the one-step model of photo-emission. The result is that, despite the sophisticated description of the photo-emission process, theory and experiment do not agree with each other, at least with regard to the spectral features of Mn-3d projected states. The deficiencies of LDA and LDA+U are related to the fact that they are basically single Slater determinant theories, and cannot describe the highly entangled many-body states corresponding to the high spin moment of the half-filled Mn-3d shell and its excited states. This is in line with our recent findings for the transition metals monoxides~\cite{thunstrom12prl109:186401}, where experimental data show a broad distribution of 3d spectral weight not describable within one-particle approximations, but well reproduced by LDA+DMFT.

\begin{figure}
\includegraphics[width=8.5cm]{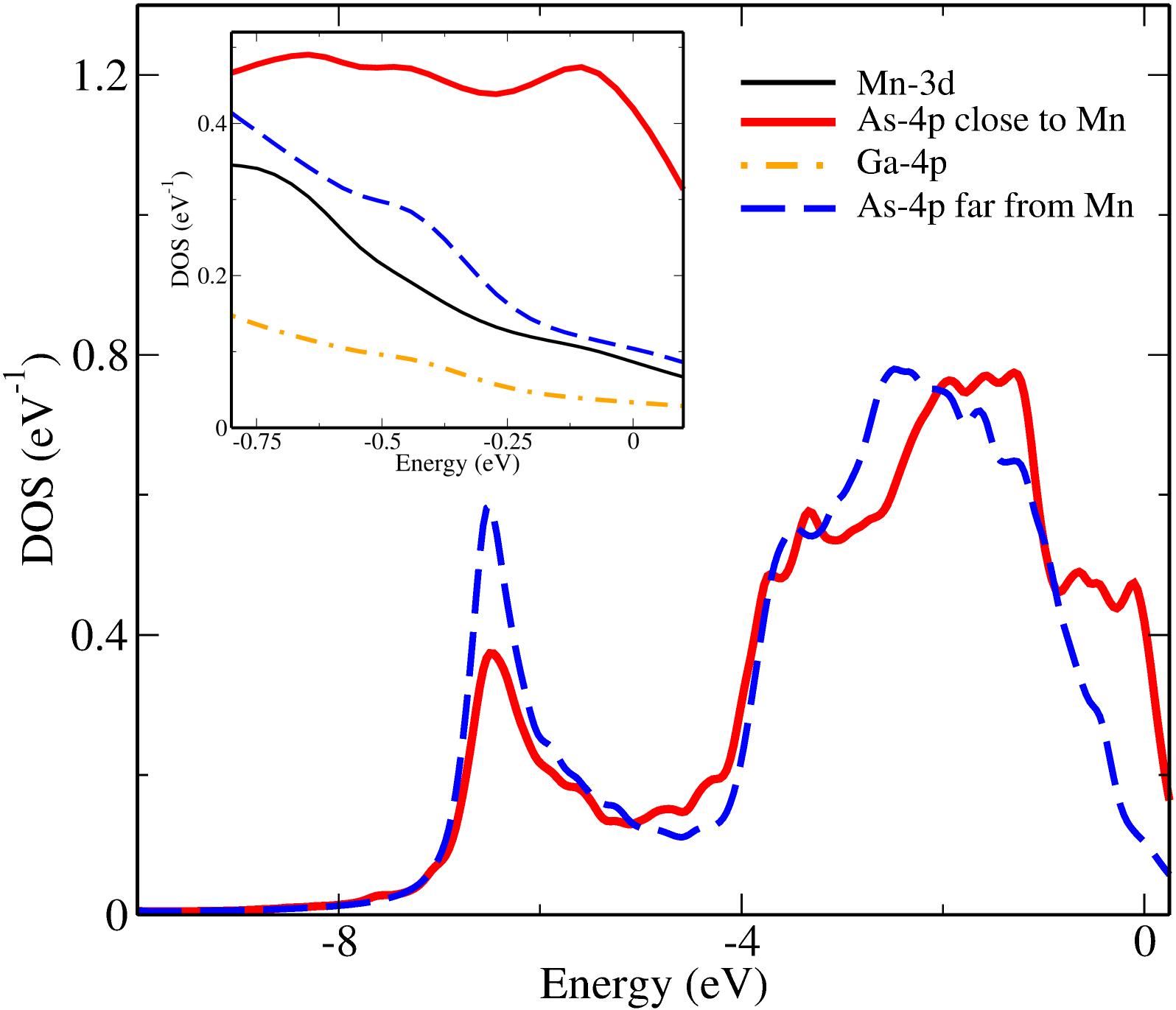}
\caption{{\bf{As-4p PDOS from LDA+DMFT.}} LDA+DMFT projected density of states of the As-4p electrons in Mn-doped GaAs, for ions that are close and far from the Mn impurity. The Fermi level is at zero energy. In the inset the projected density of states of the As-4p, the Mn-3d, and Ga-4p states are shown in the narrow region around the Fermi level.}
\label{fig:As4p_spectrum}
\end{figure}

\begin{figure}[b]
\includegraphics[width=8.5cm]{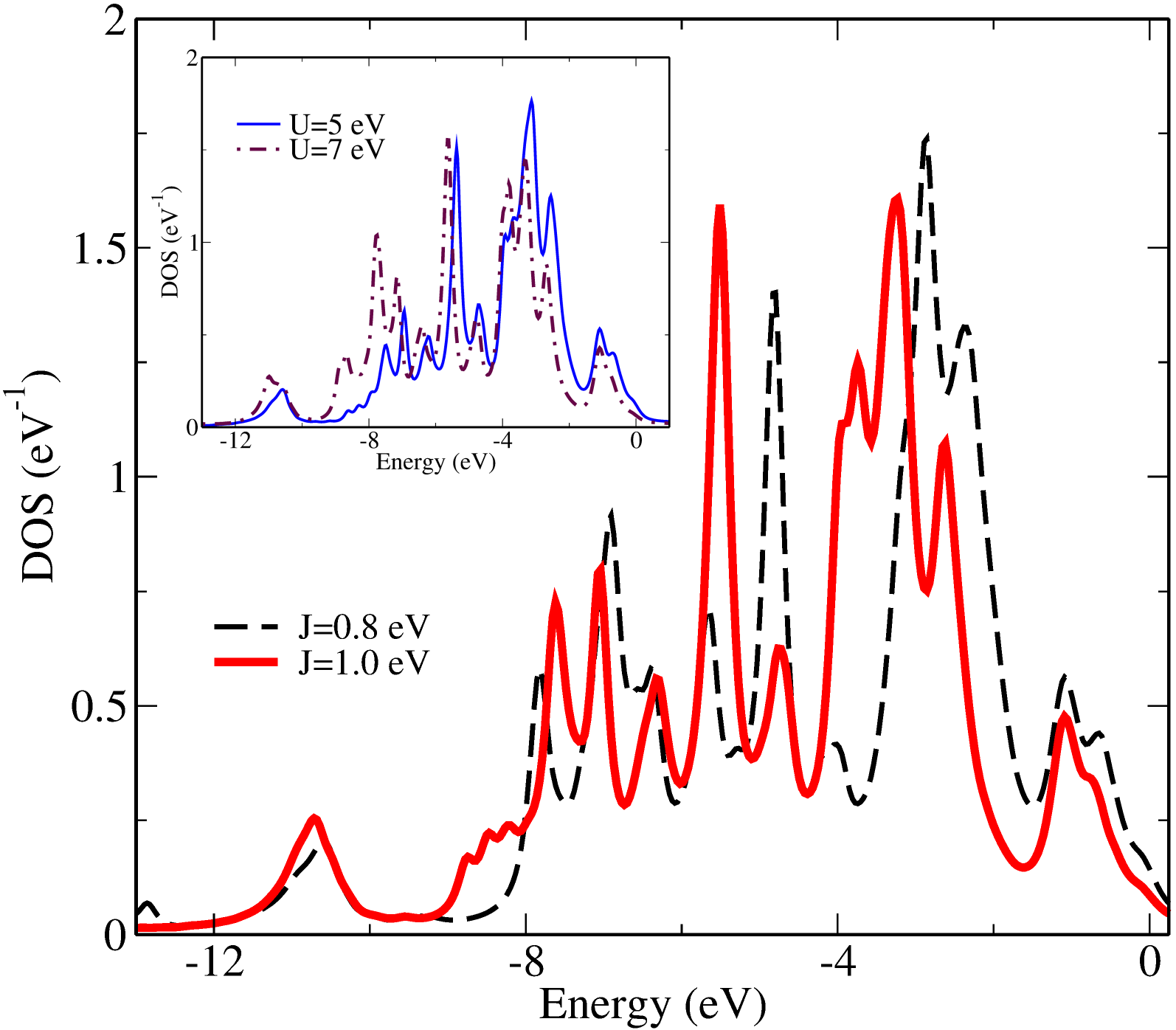}
\caption{{\bf{Mn-3d PDOS for different parametrizations of the Coulomb interaction.}} LDA+DMFT projected density of states of the Mn-3d electrons for different values of the exchange parameter $J$, and $U$ fixed to 6 eV. In the inset the same information is shown for different values of $U$, with $J$ calculated from the electron density as described in the main text. The Fermi level is at zero energy.}
\label{fig:Mn3d_UandJ}
\end{figure}

In Fig.~\ref{fig:As4p_spectrum} we report the calculated PDOS for the As-4p states, which are the main contributors to the valence band, both for an As atom far away from the Mn impurity and for an As atom close to it. As seen by inspection of the states projected on the As atom far from the Mn impurity, the Fermi level is located at the band edge of the valence band. Looking at the close region around the Fermi level (inset), one sees that the latter is mainly pinned by As-4p states, with a significant contribution of Mn-4p states (not shown). The p-states of the As atom close to the Mn atom show  hybridization with the Mn-3d states, which results in spectral features which extend above the Fermi level. As the figure shows, there is a non-negligible distortion of the As-4p states due to the presence of the impurity Mn atom, and this distortion is particularly large close to the Mn impurity. A more quantitative analysis can be made by directly evaluating the hole density for the locally projected bands in LDA+DMFT. For our supercell the Mn-3d and Mn-4p states contain respectively 1\% and 3\% of the total number of holes, while the large majority of holes has As-4p character. The first shell of As nearest neighbors (4 atoms at 0.245 nm) contains 30\% of the holes, the second shell (12 atoms at 0.469 nm) 20\%, and the third shell (6 atoms at 0.616 nm) 21\%. The last shell included in the modeling (5 atoms at 0.734 nm) exhibits a sharp drop in the hole density, containing only 7\% of the total number of holes. The first shell of Ga nearest neighbors (12 atoms at 0.4 nm) contains 9\% and 3\% for respectively the 4p and 3d states. All the other orbitals contain the remaining holes, without any significant distribution. These numbers point to a local change in the electronic structure located primarily around the impurity, and are in close agreement with a recent core level photoemission study focusing on the Ga 3d spectra from very dilute (Ga,Mn)As~\cite{kanski12jpcm24:435802}, where each Mn atom is found to affect a region corresponding to a sphere with 0.7 nm radius.
 
As shown in  Fig.~\ref{fig:Mn3d_UandJ}, the precise choice of the parameters describing the Coulomb interaction in the LDA+DMFT calculations does not lead to qualitative differences in the final spectral properties, at least if these changes are limited within a reasonable range of values. As described in the Method section, the data presented in Fig.~\ref{fig:Mn3d_spectrum} and Fig.~\ref{fig:As4p_spectrum} are obtained for an average Hubbard parameter $U = 6$ eV and for an average Hund's exchange parameter $J \simeq 1$ eV. In the inset of Fig.~\ref{fig:Mn3d_UandJ} we show the Mn-3d PDOS in LDA+DMFT for $U = 5$ and $U = 7$ eV. Even for these large changes of $U$, the main features of the two curves remain similar to what observed in Fig.~\ref{fig:Mn3d_spectrum}, and only a small redistribution of spectral weight at high binding energies can be observed. In Fig.~\ref{fig:Mn3d_UandJ}, instead, the LDA+DMFT Mn-3d PDOS shown in Fig.~\ref{fig:Mn3d_spectrum}, corresponding to $U = 6$ eV and $J \simeq 1$ eV, is compared to the LDA+DMFT data for $U = 6$ eV and $J = 0.8$ eV. The general features of the spectra are the same, but the two main peaks are shifted towards the Fermi level in the simulation with lower $J$. From these results we can see that a change of the parameters in the calculation only slightly modifies the obtained spectrum, and does not change the physical picture outlined.

\section{Discussions}
The work presented here shows experimental and theoretical photo-emission spectra of bulk Mn-doped GaAs, where good agreement between observations and theory is only found when electron correlations are treated dynamically. To be precise, dynamical mean field theory in combination with exact diagonalization is shown to reproduce the observed spectral features, whereas other impurity solvers, e.g. Hubbard-I~\cite{thunstrom09prb79:165104} or perturbative approaches~\cite{dimarco09epjb72:473}, are not adequate to this problem (data not shown). A static treatment of electron-electron repulsion, as in LDA+U, is also shown to fail when comparing with experimental spectra. This illustrates that for Mn-doped GaAs, and most probably for DMS materials in general, electronic structure calculations require a more sophisticated treatment than usually found in literature. On a more detailed level, both our theoretical and experimental results demonstrate the presence of Mn states at the top of the valence band, but that no detached impurity band forms, contrary to what is suggested in Refs.~\onlinecite{ohya10prl104:167204,dobrowolska12nm11:444}. In addition we have found that the Mn-3d states contribute very little to the carrier concentration (about 1\%). Despite the lack of a detached impurity band, our calculations reveal that the holes have a clear majority of their spectral intensity in the closest shells of As atoms around the Mn impurity, a fact to be taken into consideration when analyzing the exchange mechanism of this material. Our results are in very good agreement with recent hard X-ray photoemission data, which, differently from our study, are obtained by probing directly the ferromagnetic phase~\cite{gray12natmat11:957}. However, the sophisticated photo-emission theory used in that work is still based on the LDA approximation, and therefore suffers of all the inaccuracies mentioned above. Finally, in our theoretical framework, a more quantitative and definite conclusion on the magnetism could likely be obtained by approaching the problem of the evolution of the ordering temperature for different Mn concentrations from first principles.

\section{Methods}
\subsection{Experiment}
The photo-emission data were obtained at the Swedish synchrotron radiation facility MAX-lab using the surface end station of the I511 undulator beamline~\cite{denecke99jesrp101:971}. The samples were prepared in a local molecular beam epitaxy (MBE) system, and were transferred to the photo-emission station in a portable UHV chamber without being exposed to atmosphere. The Mn concentrations were determined during growth by means of reflection high-energy electron diffraction (RHEED) oscillations, as described earlier~\cite{sadowski00jvstb18:1697}. Survey spectra recorded after transfer showed contamination-free surfaces, and low-energy electron diffraction (LEED) showed (1x2) surface reconstruction. For very low Mn concentration (0.1\%) the fractional spots of GaAs(100) c(4x4) were still present, but clearly stretched along the $\langle100\rangle$ azimuths, reflecting a transition for (1x2) pattern. All spectra presented here were obtained at room temperature from as-grown samples, i.e. samples not subjected to post-growth annealing. After the photo-emission experiment the magnetic properties were measured ex situ in a SQUID setup. The sample with 6\% Mn showed ferromagnetic behavior below 55 K, while none of the other samples showed long range order above 5 K.

Unlike most previous experimental works (e.g. Refs.~\onlinecite{sadowski00jvstb18:1697,okabayashi98prb58:R4211,rader09pssb246:1435,okabayashi01prb64:125304}) the present experiments were carried out on as-grown samples, transferred in ultrahigh vacuum between the growth- and analysis systems. This is an essential difference, since the investigated (Ga,Mn)As material is known to be metastable and its surface cannot be reliably restored by traditional preparation methods like Ar sputtering and annealing. Indeed according to our previous studies~\cite{asklund02prb66:115319}, the main Mn-3d emission valence band structure was found significantly closer to the valence band maximum than that reported in all above mentioned reports, and a shift towards higher binding energy was demonstrated to be induced by post-growth annealing~\cite{adell04prb70:125204}. While it can be argued that the changes were observed at temperatures above 300$^\circ$C, i.e. higher than the annealing temperatures applied in the post-Ar-sputtering treatments, the energy deposited by the Ar sputtering itself is certainly sufficient to induce structural modifications. By exploiting the momentum selectivity in resonant photo-emission we can focus specifically on the 3d states. As will be shown,
the 3d states are even shallower than suggested by spectra probing the integrated valence band, and the deviation from the value reported in Refs. \onlinecite{okabayashi98prb58:R4211,rader09pssb246:1435} is even larger. 

Valence band spectra were recorded at two photon energies, one on the 2p$_{3/2}$ absorption peak (A in the inset of Fig.~\ref{fig:Mn_exp_spectrum}), the other 2 eV below the peak (B in the inset of Fig.~\ref{fig:Mn_exp_spectrum}), and the Mn-3d resonant emission was obtained from the difference between the two spectra. In a previous study \cite{rader04prb69:075202} corresponding structures were extracted by combining spectra from (Ga,Mn)As and clean GaAs obtained at the same excitation energy. The main advantage of the procedure adopted here is that it avoids artifacts that inevitably arise due to different surface states on different surfaces. In the extraction of difference spectra it is of course important, particularly in the most dilute cases, that the spectra are properly normalized and aligned in energy. This was accomplished by including the Ga-3d peak in the recorded energy range. By means of this peak the spectra could be aligned with a precision of 1 meV, and the normalization was checked via the absence of any systematic structures in the energy region of Ga-3d emission. For the most dilute cases it turned out that thermal effects, mainly associated with gradual monochromator heating had to be taken into account. To minimize these effects, the recording time for each pair of spectra was reduced to about 2 minutes. To achieve reasonable statistics, a number of difference spectra (typically 50-100) were added after proper alignment. In contrast to the recently published hard X-ray photo-emission from (Ga,Mn)As~\cite{gray12natmat11:957}, the present data are quite surface sensitive due to the relatively short electron mean free path (around 1.5 nm). However, due to the rapidly falling Mn-3d/As-4p cross section ratio with increasing photon energy~\cite{scofield73ucrl:51326} the emission from Mn-3d in Ref.~\onlinecite{gray12natmat11:957} is hardly discernible. In contrast, under the resonant conditions used here the Mn 3d emission is enhanced selectively and dominates the spectrum for samples with Mn concentrations above 1\%.

\subsection{Theory}
The theoretical spectra have been calculated by means of a LDA+DMFT technique~\cite{dimarco09prb79:115111,granas12cms55:295} based on a full-potential linear-muffin tin orbital method (FP-LMTO)~\cite{RSPt_book}, which is particularly suitable for large supercell geometries. The Mn-doped GaAs was modeled with a supercell Ga$_{26}$MnAs$_{27}$, corresponding to a Mn concentration of 3.7\%. The experimental lattice constant $a=10.681 \text{ a.u.}$ was used, and the Brillouin zone was sampled with a mesh of 256 {\bf{k}}-points.
The LDA+DMFT scheme was applied to explicitly treat the local Coulomb interaction between the localized Mn-3d electrons. The 4-index rotationally-invariant Coulomb interaction matrix was generated from the Slater parameters $F^0$, $F^2$ and $F^4$. The choice of the average Coulomb repulsion $F^0$, which corresponds to the Hubbard U, is rather problematic, since no calculations based on constrained LDA or RPA methods are found in literature. Therefore we have considered values between 4 eV and 7 eV, which are the accepted strengths of the Coulomb repulsion for bulk metallic $\gamma$-Mn~\cite{dimarco09epjb72:473} and MnO~\cite{thunstrom12prl109:186401}. The main results of the paper are presented for the intermediate value $U = 6$ eV, while results for smaller and larger values are discussed at the end of the Results section. $F^2$ and $F^4$ are easier to evaluate, and therefore were calculated directly from the electronic density as done in Ref.~\onlinecite{thunstrom12prl109:186401}. The calculated values correspond to the average Hund's exchange parameter $J \simeq 1$ eV. The LDA+DMFT results for $J = 0.8$ eV, shown in Fig.~\ref{fig:Mn3d_UandJ}, were based on $F^2$ and $F^4$ obtained by means of fixed atomic ratios~\cite{kotliar06rmp78:865,held07ap56:829}.

The effective impurity problem arising in LDA+DMFT has been solved through exact diagonalisation (ED) method, as described in Ref.~\onlinecite{thunstrom12prl109:186401}. The fermionic bath interacting with the atomic impurity has been approximated by means of 22 auxiliary bath spin-orbitals: 18 bath states were coupled to the strongly hybridizing Mn t$_{2g}$ states, while only 4 bath states were coupled to the weakly hybridizing Mn e$_g$ states. All the calculations have been made for the paramagnetic phase at $\text{T}=400 \text{ K}$ and using 1200 fermionic Matsubara frequencies, and double counting problem has been considered in the fully localized limit (FLL) \cite{kotliar06rmp78:865,held07ap56:829}. Finally in the calculation of the hole densities we have considered only the atoms included in one supercell and their multiplicity, consistently with our physical modeling and previous literature~\cite{sandratskii04prb69:195203}.

\section{Acknowledgment}
Financial support from the Swedish Research Council (VR) and Energimyndigheten (STEM), is acknowledged. Calculations have been performed at the Swedish national computer centers UPPMAX, PDC, HPC2N and NSC. O.E. is also grateful to the European Research Council (grant 247062 - ASD) and the Knut and Alice Wallenberg Foundation, for financial support.

\section{Contributions}
O.E., M.I.K., and J.K. planned the project. J.S. prepared the samples for the experiment. J.K. carried out the experiment. J.K. and K.K. performed the data analysis. S.L. made a preliminary computational study. I.D.M. and P.T. developed the code for the simulations. I.D.M. made all the calculations. I.D.M., P.T., and O.E. wrote the paper. All co-authors contributed to discuss the results, revise the manuscript, and reply to Referees' criticisms.

\section{Additional information}
{\bf{Competing financial interests:}} The authors declare no competing financial interests. 




%

\end{document}